\documentclass[final,english]{bullsrsl}[2022/06/15]

\usepackage[latin1]{inputenc}
\usepackage[T1]{fontenc}

\usepackage{natbib} 
\usepackage{graphicx}

\begin{document}
\title{X-raying the wind-wind collisions in HD\,168112 and HD\,167971}

\author[affil={1}, corresponding]{Gregor}{Rauw}
\author[affil={2}]{Ronny}{Blomme}
\author[affil={1}]{Ya\"el}{Naz\'e}
\author[affil={3}]{Delia}{Volpi}
\affiliation[1]{STAR Institute, Universit\'e de Li\`ege, All\'ee du 6 Ao\^ut, 19c, B\^at B5c, 4000 Li\`ege, Belgium}
\affiliation[2]{Royal Observatory of Belgium, Ringlaan 3, 1180 Brussels, Belgium}
\affiliation[3]{FSM, Universit\'e Libre de Bruxelles, Campus Erasme, 1070 Anderlecht, Belgium}
\correspondance{g.rauw@uliege.be}
\date{29 May 2024}
\maketitle

\begin{abstract}
The O-type long-period binary HD\,168112 and triple HD\,167971 star systems have been known for several decades for their non-thermal synchrotron radio emission. This emission arises from relativistic electrons accelerated in the hydrodynamic shocks of the wind collisions in these systems. Such wind collisions are expected to produce a strong X-ray emission that varies as a function of orbital phase. In wide eccentric binaries, such as our targets, the X-ray emission arises from an adiabatic plasma and its intensity should scale as the inverse of the orbital separation. We present a set of {\it XMM-Newton} observations of these systems which help us gain insight into the properties of their wind interactions.
\end{abstract}


\begin{altabstract}
  \textbf{Analyse en rayons X des collisions de vents stellaires dans HD\,168112 et HD\,167971}.
  La binaire O+O \`a p\'eriode longue HD\,168112 et le syst\`eme triple HD\,167971 sont connus depuis les ann\'ees 1980 pour leur \'emission radio synchrotron. Cette \'emission radio est produite par des \'electrons relativistes acc\'el\'er\'es par les chocs hydrodynamiques de la collision des vents stellaires dans ces syst\`emes. Ces collisions peuvent \'egalement g\'en\'erer une intense \'emission en rayons X qui varie en fonction de la phase orbitale. Dans des binaires excentriques \`a longue p\'eriode orbitale, comme celles \'etudi\'ees ici, le plasma en aval du choc se trouve dans le r\'egime adiabatique et l'intensit\'e de l'\'emission X devrait varier comme l'inverse de la s\'eparation orbitale. Ici, nous discutons des observations {\it XMM-Newton} de HD\,168112 et HD\,167971 qui nous permettent de mieux comprendre les propri\'et\'es de collisions de vents au sein de ces syst\`emes.
\end{altabstract}

\altkeywords{\'etoiles massives, \'etoiles individuelles (HD\,168112; HD\,167971), binaires s\'err\'ees, \'emission de rayons X stellaires}

\section{Introduction}
HD\,168112 and HD\,167971 are known for their non-thermal radio emission hinting at energetic wind-wind collisions \citep[][and references therein]{Blo05,Blo07}. HD\,167971 is a hierarchical triple system consisting of a 3.32-day O4/5\,If + O4/5\,V-III eclipsing binary with an O8\,Iaf tertiary on a 21.2\,yrs orbit with an eccentricity of $e=0.53$ \citep[][and references therein]{Iba13,LeB17,Mai19}. HD\,168112 is an eccentric ($e=0.75$) binary consisting of an O4.5\,IV((f)) primary and an O5.5\,V(n)((f)) secondary on a 514-day orbit \citep{Put23,Blo24}.
\begin{figure}[htb]
\centering
\resizebox{12cm}{!}{\includegraphics{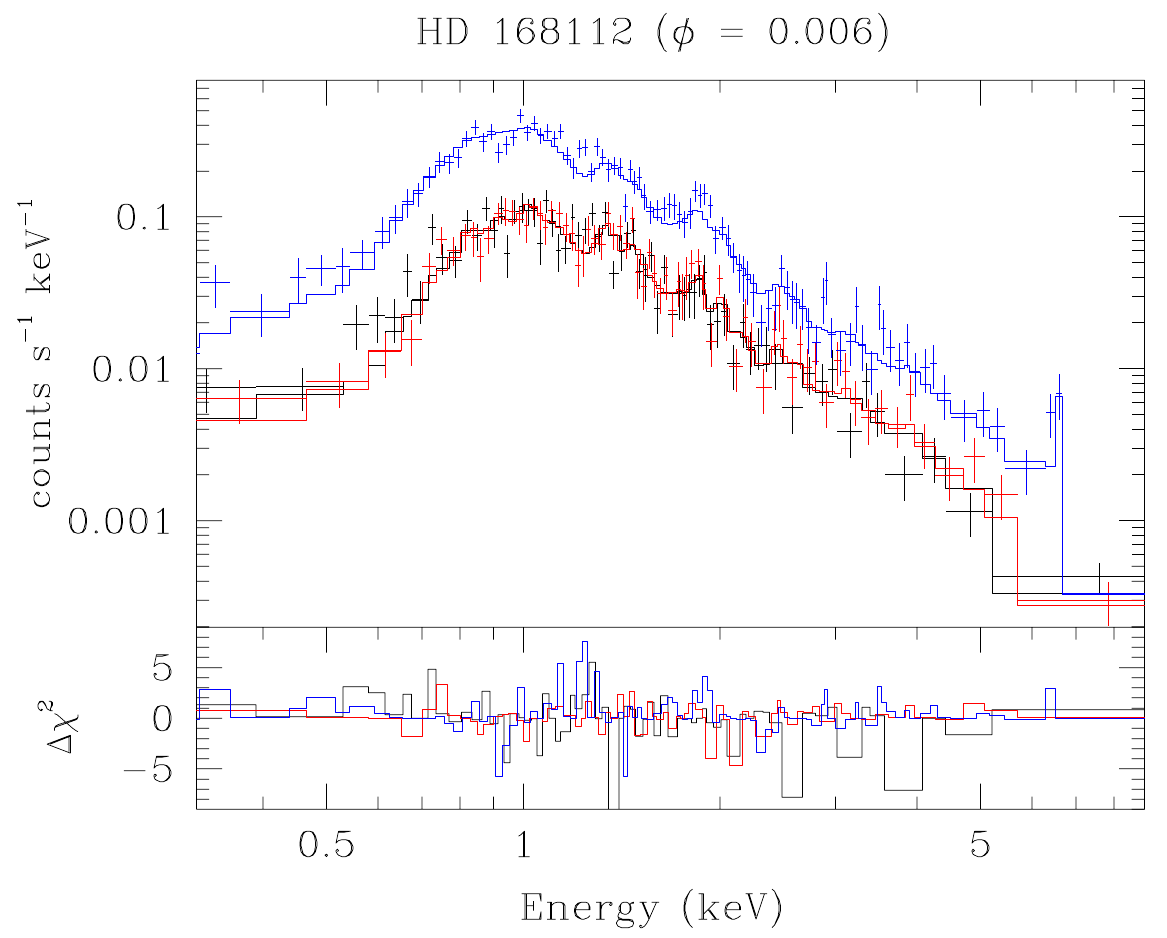}}
\bigskip
\begin{minipage}{12cm}
  \caption{{\it XMM-Newton} EPIC spectra of HD\,168112 near periastron passage in March 2023. In the upper panel, the observed data are shown with their error bars, whereas the best-fit 3-T model (in {\tt xspec} terminology ${\tt TBabs}_{\rm ISM} * {\tt phabs}_{\rm wind} * \sum_{i=1}^3 {\tt apec}(kT_i)$) is illustrated by the histogram. Black, red and blue colours correspond to EPIC-MOS1, EPIC-MOS2 and EPIC-pn data, respectively. The EPIC-pn spectrum reveals an Fe {\sc xxv} emission line at 6.7\,keV, clearly demonstrating the thermal nature of the plasma. The lower panel illustrates the contribution $\Delta\chi^2$ for each energy bin to the global $\chi^2$ of the fit, multiplied by the sign of the difference between observation and model.\label{spec168112}}
\end{minipage}
\end{figure}

HD\,168112 and HD\,167971 are members of the NGC\,6604 open cluster and fall into a single field of view of the {\it XMM-Newton} EPIC instrument. A total of six observations of NGC\,6604 were obtained with {\it XMM-Newton}. The first two were taken in 2002, a third one was obtained in 2014 and three additional observations were collected around the March 2023 periastron passage of HD\,168112. Spectra and light curves were extracted. The X-ray spectra of both sources were analysed with the {\tt xspec} software \citep{Arn96} and are well represented by multi-temperature (2-T and 3-T) optically thin thermal plasma models (Fig.\,\ref{spec168112}). We used these models to derive the X-ray fluxes in the 0.5 - 10.0\,keV energy domain corrected for the absorption by the interstellar medium. Whilst the light curves do not reveal any significant intra-pointing variations, we find strong variations of the fluxes between the observations.

\section{HD\,168112}
\begin{figure}[htb]
\centering
\resizebox{12cm}{!}{\includegraphics{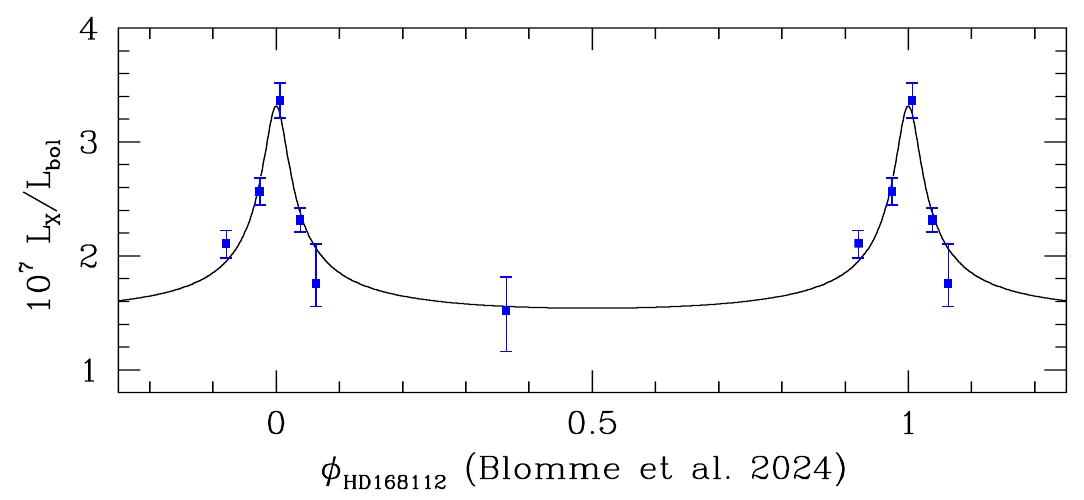}}
\bigskip
\begin{minipage}{12cm}
  \caption{$L_{\rm X}/L_{\rm bol}$ ratio of HD\,168112 as a function of orbital phase \citep[ephemerides of][]{Blo24}. The X-ray luminosities were computed from the 0.5 - 10\,keV fluxes corrected for interstellar absorption and adopting a distance of 2\,kpc, in agreement with the {\it Gaia} parallax. The bolometric luminosities were taken from \citet{Put23}. The solid line yields the best-fit relation: $L_{\rm X}/L_{\rm bol} (10^{-7}) = 0.51\,a/r(\phi) + 1.25$.\label{fx168112}} 
\end{minipage}
\end{figure}

For this system, we observe a strong increase of the X-ray flux at periastron (Fig.\,\ref{fx168112}). The flux variations are well-described by a $a/r(\phi)$ relation where $r(\phi)$ is the instantaneous orbital separation between the stars at phase $\phi$ and $a$ is the semi-major axis of the orbit (Fig.\,\ref{fx168112}). This result agrees with the expectation for a wind interaction in the adiabatic regime \citep{SBP}. We observe no deviations from a $a/r(\phi)$ behaviour. Such deviations would be expected if shock modification due to relativistic electrons were important. Furthermore, the shock remains stable around periastron passage \citep[unlike in some other eccentric systems such as WR\,21a,][]{Gos16}. The part of the X-ray emission that remains constant with orbital phase is due to the wind-embedded shocks of both stars and follows $L_{\rm X}/L_{\rm bol} = 1.25 \times 10^{-7}$ in agreement with the canonical relation for O-type stars \citep[e.g.,][]{Naz09}.
\section{HD\,167971}
\begin{figure}[htb]
\centering
\resizebox{12cm}{!}{\includegraphics{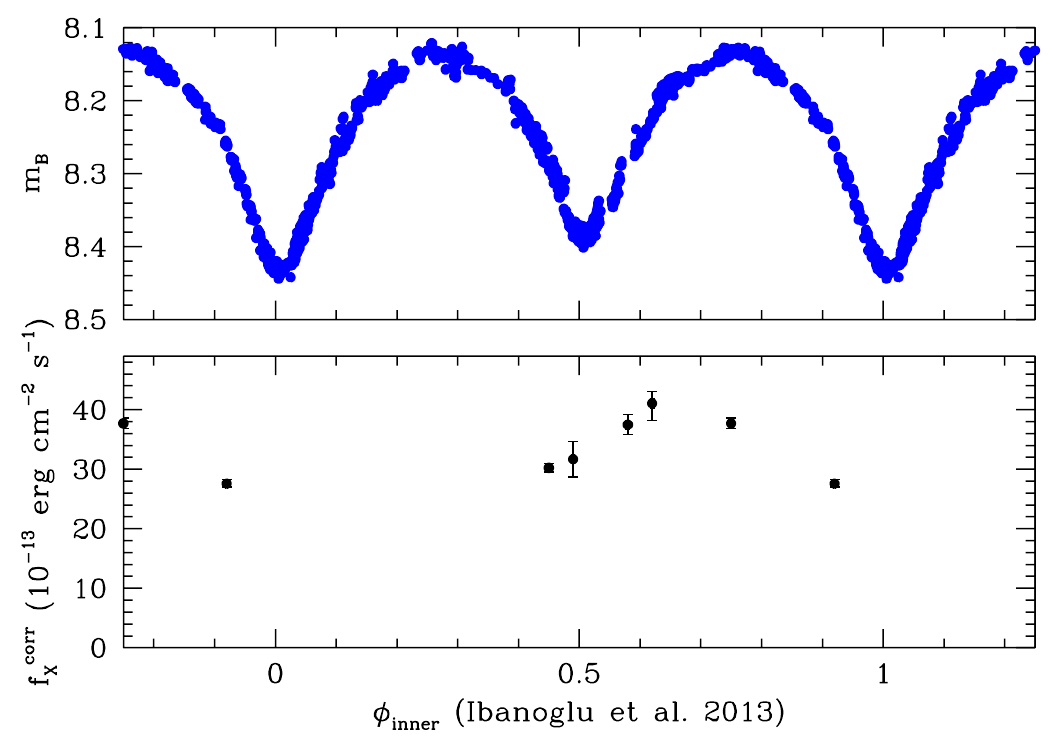}}
\bigskip
\begin{minipage}{12cm}
  \caption{Optical and X-ray variations of HD\,167971 folded with the ephemerides of the inner binary's orbit \citep{Iba13}. The top panel illustrates $B$-band photometry from \citet{May10}, whilst the bottom panel displays the 0.5 - 10\,keV fluxes corrected for interstellar absorption.\label{fxMYSer}} 
\end{minipage}
\end{figure}

The mean $L_{\rm X}/L_{\rm bol}$ ratio of this system amounts to $4.8 \times 10^{-7}$. The total X-ray emission consists of various contributions arising from (i) the intrinsic shocks embedded in the winds of the individual components of the triple system, (ii) the wind-wind interaction in the inner binary, and (iii) the collision of the wind of the inner binary with the wind of the tertiary star.  
Unfortunately, the existing data do not provide a sufficient sampling of the outer (21.2\,yrs) orbital cycle to investigate the variations in contribution (iii) to the total X-ray emission. We nevertheless observe strong variations of the X-ray flux by $\sim 40$\%, peak to peak. These variations occur on timescales consistent with the 3.32 days period of the eclipsing binary. This suggests that the radiative wind interaction zone in the inner binary contributes a significant fraction of the overall X-ray emission. Under these circumstances, the most likely cause of the observed variability would be occultation of the wind-wind interaction zone by the bodies of the stars in the inner binary possibly combined with changes of the circumstellar column density along our sightline. The observed shift in orbital phase between the variations of the X-rays and the optical light curve could then be the consequence of the Coriolis deflection (see Fig.\,\ref{fxMYSer}).


\begin{furtherinformation}
\begin{orcids}
  \orcid{0000-0003-4715-9871}{Gregor}{Rauw}
  \orcid{0000-0002-2526-346X}{Ronny}{Blomme}
  \orcid{0000-0003-4071-9346}{Ya\"el}{Naz\'e}
  \orcid{0009-0001-1183-0021}{Delia}{Volpi}  
\end{orcids}

\begin{authorcontributions}
  GR reduced the X-ray data and led their analysis. RB reduced the radio data and provided the orbital solution of HD\,168112. All authors contributed to the interpretation of the results. 
\end{authorcontributions}

\begin{conflictsofinterest}
The authors declare no conflict of interest.
\end{conflictsofinterest}

\end{furtherinformation}

\bibliographystyle{bullsrsl-en}

\bibliography{G_Rauw_1}

\begin{thebibliography}{12}
\providecommand{\natexlab}[1]{#1}
\providecommand{\url}[1]{#1}
\providecommand{\urlprefix}{URL }

\bibitem[{{Arnaud}(1996)}]{Arn96}
{Arnaud}, K.~A. (1996) {XSPEC: The First Ten Years}.
\newblock In Astronomical Data Analysis Software and Systems V, edited by
  {Jacoby}, G.~H. and {Barnes}, J., vol. 101 of \emph{Astronomical Society of
  the Pacific Conference Series}, p.~17.

\bibitem[{{Blomme} et~al.(2007){Blomme}, {De Becker}, {Runacres}, {van Loo} and
  {Setia Gunawan}}]{Blo07}
{Blomme}, R., {De Becker}, M., {Runacres}, M.~C., {van Loo}, S. and {Setia
  Gunawan}, D.~Y.~A. (2007) {Non-thermal radio emission from O-type stars. II.
  HD 167971}.
\newblock A\&A, 464(2), 701--708.
\newblock \url{https://doi.org/10.1051/0004-6361:20054602}.

\bibitem[{{Blomme} et~al.(2024){Blomme}, {Rauw}, {Volpi}, {Naz{\'e}} and
  {Abdul-Masih}}]{Blo24}
{Blomme}, R., {Rauw}, G., {Volpi}, D., {Naz{\'e}}, Y. and {Abdul-Masih}, M.
  (2024) {The colliding-wind binary HD 168112}.
\newblock A\&A, in press.
\newblock \url{https://doi.org/10.48550/arXiv.2405.03247}.

\bibitem[{{Blomme} et~al.(2005){Blomme}, {van Loo}, {De Becker}, {Rauw},
  {Runacres}, {Setia Gunawan} and {Chapman}}]{Blo05}
{Blomme}, R., {van Loo}, S., {De Becker}, M., {Rauw}, G., {Runacres}, M.~C.,
  {Setia Gunawan}, D.~Y.~A. and {Chapman}, J.~M. (2005) {Non-thermal radio
  emission from O-type stars. I. HD168112}.
\newblock A\&A, 436(3), 1033--1040.
\newblock \url{https://doi.org/10.1051/0004-6361:20042383}.

\bibitem[{{Gosset} and {Naz{\'e}}(2016)}]{Gos16}
{Gosset}, E. and {Naz{\'e}}, Y. (2016) {The X-ray light curve of the massive
  colliding wind Wolf-Rayet + O binary WR 21a}.
\newblock A\&A, 590, A113.
\newblock \url{https://doi.org/10.1051/0004-6361/201527051}.

\bibitem[{{Ibanoglu} et~al.(2013){Ibanoglu}, {{\c{C}}ak{\i}rl{\i}} and
  {Sipahi}}]{Iba13}
{Ibanoglu}, C., {{\c{C}}ak{\i}rl{\i}}, {\"O}. and {Sipahi}, E. (2013) {MY
  Serpentis: a high-mass triple system in the Ser OB2 association}.
\newblock MNRAS, 436(1), 750--758.
\newblock \url{https://doi.org/10.1093/mnras/stt1616}.

\bibitem[{{Le Bouquin} et~al.(2017){Le Bouquin}, {Sana}, {Gosset}, {De Becker},
  {Duvert}, {Absil}, {Anthonioz}, {Berger}, {Ertel}, {Grellmann}, {Guieu},
  {Kervella}, {Rabus} and {Willson}}]{LeB17}
{Le Bouquin}, J.~B., {Sana}, H., {Gosset}, E., {De Becker}, M., {Duvert}, G.,
  {Absil}, O., {Anthonioz}, F., {Berger}, J.~P., {Ertel}, S., {Grellmann}, R.,
  {Guieu}, S., {Kervella}, P., {Rabus}, M. and {Willson}, M. (2017) {Resolved
  astrometric orbits of ten O-type binaries}.
\newblock A\&A, 601, A34.
\newblock \url{https://doi.org/10.1051/0004-6361/201629260}.

\bibitem[{{Ma{\'\i}z Apell{\'a}niz} et~al.(2019){Ma{\'\i}z Apell{\'a}niz},
  {Trigueros P{\'a}ez}, {Negueruela}, {Barb{\'a}}, {Sim{\'o}n-D{\'\i}az},
  {Lorenzo}, {Sota}, {Gamen}, {Fari{\~n}a}, {Salas}, {Caballero}, {Morrell},
  {Pellerin}, {Alfaro}, {Herrero}, {Arias} and {Marco}}]{Mai19}
{Ma{\'\i}z Apell{\'a}niz}, J., {Trigueros P{\'a}ez}, E., {Negueruela}, I.,
  {Barb{\'a}}, R.~H., {Sim{\'o}n-D{\'\i}az}, S., {Lorenzo}, J., {Sota}, A.,
  {Gamen}, R.~C., {Fari{\~n}a}, C., {Salas}, J., {Caballero}, J.~A., {Morrell},
  N.~I., {Pellerin}, A., {Alfaro}, E.~J., {Herrero}, A., {Arias}, J.~I. and
  {Marco}, A. (2019) {MONOS: Multiplicity Of Northern O-type Spectroscopic
  systems. I. Project description and spectral classifications and visual
  multiplicity of previously known objects}.
\newblock A\&A, 626, A20.
\newblock \url{https://doi.org/10.1051/0004-6361/201935359}.

\bibitem[{{Mayer} et~al.(2010){Mayer}, {Bo{\v{z}}i{\'c}}, {Lorenz} and
  {Drechsel}}]{May10}
{Mayer}, P., {Bo{\v{z}}i{\'c}}, H., {Lorenz}, R. and {Drechsel}, H. (2010)
  {Sixty four nights of U BV photometry of early-type stars at La Silla}.
\newblock Astronomische Nachrichten, 331(3), 274.
\newblock \url{https://doi.org/10.1002/asna.200911303}.

\bibitem[{{Naz{\'e}}(2009)}]{Naz09}
{Naz{\'e}}, Y. (2009) {Hot stars observed by XMM-Newton. I. The catalog and the
  properties of OB stars}.
\newblock A\&A, 506(2), 1055--1064.
\newblock \url{https://doi.org/10.1051/0004-6361/200912659}.

\bibitem[{{Putkuri} et~al.(2023){Putkuri}, {Gamen}, {Morrell}, {Ma{\'\i}z
  Apell{\'a}niz}, {Arias}, {Sim{\'o}n-D{\'\i}az}, {Ferrero}, {Rodr{\'\i}guez},
  {Sota}, {Benvenuto} and {Barb{\'a}}}]{Put23}
{Putkuri}, C., {Gamen}, R., {Morrell}, N.~I., {Ma{\'\i}z Apell{\'a}niz}, J.,
  {Arias}, J.~I., {Sim{\'o}n-D{\'\i}az}, S., {Ferrero}, G.~A.,
  {Rodr{\'\i}guez}, C.~N., {Sota}, A., {Benvenuto}, O.~G. and {Barb{\'a}},
  R.~H. (2023) {The spectroscopic orbit of HD 168112 A,B in NGC 6604: another
  massive binary target for interferometry}.
\newblock MNRAS, 525(4), 6084--6096.
\newblock \url{https://doi.org/10.1093/mnras/stad2657}.

\bibitem[{{Stevens} et~al.(1992){Stevens}, {Blondin} and {Pollock}}]{SBP}
{Stevens}, I.~R., {Blondin}, J.~M. and {Pollock}, A.~M.~T. (1992) {Colliding
  Winds from Early-Type Stars in Binary Systems}.
\newblock ApJ, 386, 265.
\newblock \url{https://doi.org/10.1086/171013}.

\end{thebibliography}

\end{document}